\documentclass[aps,prc,letterpaper,11pt,twoside,tightenlines,nofootinbib,showpacs,preprint]{revtex4-1}
\usepackage{graphicx}
\usepackage[sort&compress]{natbib}
\usepackage{subfigure}
\usepackage{amsmath}
\usepackage{amsfonts}
\usepackage{cancel}
\usepackage{amssymb}

\begin{document}
\arraycolsep1.5pt
\newcommand{\Ima}{\textrm{Im}}
\newcommand{\Rea}{\textrm{Re}}
\newcommand{\mev}{\textrm{ MeV}}
\newcommand{\be}{\begin{equation}}
\newcommand{\ee}{\end{equation}}
\newcommand{\ba}{\begin{eqnarray}}
\newcommand{\ea}{\end{eqnarray}}
\newcommand{\gev}{\textrm{ GeV}}
\newcommand{\nn}{{\nonumber}}
\newcommand{\dtres}{d^{\hspace{0.1mm} 3}\hspace{-0.5mm}}
\newcommand{\rts}{ \sqrt s}
\newcommand{\non}{\nonumber \\[2mm]}

\title{\boldmath The ``$a_1(1420)$'' peak as the $\pi f_0(980)$ decay mode of the $a_1(1260)$.}

\author{F. Aceti$^{1}$, L. R. Dai$^{1,2}$ and E. Oset$^{1}$}
\affiliation{$^{1}$Departamento de F\'{\i}sica Te\'orica, Universidad de Valencia and IFIC, Centro Mixto Universidad de
Valencia-CSIC, Institutos de Investigaci\'on de Paterna, Apartado. 22085, 46071 Valencia,
Spain\\ \\
$^{2}$Department of Physics, Liaoning Normal University, Dalian 116029, China,\\
 }

\date{\today}

\begin{abstract}
We  study the decay mode of the $a_1(1260)$ into a $\pi^+$ in p-wave and the $f_0(980)$ that decays into $\pi^+ \pi^-$ in s-wave. The mechanisms proceeds via a triangular mechanism where the $a_1(1260)$ decays into $K^* \bar K$, the $K^*$ decays to an external $\pi^+$ and an internal $K$ that fuses with the $\bar K$ to produce the $f_0(980)$ resonance. The mechanism develops a singularity at a mass of the $a_1(1260)$ around 1420 MeV, producing a peak in the cross section of the $\pi p$ reaction, used to generate the mesonic final state, which provides a natural explanation of all the features observed in the COMPASS experiment, where a peak observed at this energy is tentatively associated to a new resonance called $a_1(1420)$. On the other hand, the triangular singularity studied here gives rise to a remarkable feature, where a peak is seen for a certain decay channel of a resonance at an energy about 200 MeV higher than its nominal mass. 
\end{abstract}
\pacs{11.80.Gw, 12.38.Gc, 12.39.Fe, 13.75.Lb}

\maketitle
\raggedbottom

\section{Introduction}
\label{Intro}
The COMPASS collaboration reported the observation of a peak around 1420 MeV in the $f_0(980) \pi$ final state, with the pion in P-wave. Then, the $f_0(980)$ decays into $\pi^+ \pi^-$ in S-wave. This state was observed in the diffractive scattering of 190 GeV $\pi^-$ beam on a proton target \cite{Adolph:2015pws} and was claimed as a signal of a new resonance that was named the $"a_1(1420)"$ resonance, since the quantum numbers of the final state correspond to a $I^G(J^{PC})=1^-(1^{++})$ configuration. The signal was also observed in the VES experiment \cite{Khokhlov:2014nha} in the channel $\pi^- \pi^0 \pi^0$.

The triangular singularity played an important role describing the $\eta(1405)\to\pi^0 a_0(980)$ and the isospin forbidden $\eta(1405)\to \pi^0 f_0(980)$ decays measured in \cite{BESIII:2012aa} and studied theoretically in \cite{Wu:2011yx,Wu:2012pg}. A reevaluation of the triangular diagrams was done in \cite{Aceti:2012dj} providing a natural way to regularize the loops in the $\eta(1405)\to\pi^0 a_0(980)$ decay and giving a good description of the invariant mass distributions for the process. A search for related singularities in other physical processes has been done in \cite{Liu:2015taa}. 

Recently a paper \cite{Ketzer:2015tqa} challenged the claim of the COMPASS peak as a signal of a new resonance, providing a natural explanation of it based on the triangular singularity that unavoidably stems from the decay of the $a_1(1260)$ resonance into $K^* \bar K$ followed by $K^* \to \pi K$, with the pion emitted and the remaining $K \bar K$ merging into the $f_0(980)$ resonance. It was also noted in \cite{Wang:2015cis} that, if the $a_1(1420)$ structure arises from the $K\bar K^*$ decay mode of the $a_1(1260)$, then the production rates of the $a_1(1420)$ in $B$ decays are totally determined by the rates for the $a_1(1260)$. 

The triangle singularities were studied by Landau \cite{Landau:1959fi} and they appear from processes involving a Feynman diagram which has a loop with three intermediate particles, when the three of them are placed on shell and the momenta of the particles are collinear. Recently \cite{Guo:2015umn,Liu:2015fea,Guo:2016bkl}, the triangle singularities have also been advocated as an explanation for the observed peak in the $J/\psi p$ spectrum in the $\Lambda_b \to J/\psi K^- p$ \cite{Aaij:2015tga,Aaij:2014zoa} and $\Lambda_b \to J/\psi \pi^- p$ \cite{Aaij:2016ymb} decays, from where the existence of two pentaquarks states has been claimed (see recent talks of S. Stone at the Blois meeting \cite{sheldon} and T. Skwarnicki at the Meson Conference \cite{tomasz} on the latter reaction).  There is, however, a very important difference between the work of \cite{Ketzer:2015tqa} and those of \cite{Guo:2015umn,Liu:2015fea,Guo:2016bkl}, since in the latter works one does not have information on the couplings needed and then one does not have any idea on the strength of the singularity. On the contrary, in the work of \cite{Ketzer:2015tqa} the dynamics is well known and then a clear prediction of the strength of the singularity can be made. Thus, while the works of \cite{Guo:2015umn,Liu:2015fea,Guo:2016bkl} provide a speculation, the one of \cite{Ketzer:2015tqa} is more than that, since, if confirmed independently, then it offers a natural explanation for all the experimental facts observed in \cite{Adolph:2015pws}.

  The purpose of the present work is to provide an independent confirmation of the results and conclusions of \cite{Ketzer:2015tqa}. At the same time we offer a technically different derivation and we can provide an answer to questions which were left open in \cite{Adolph:2015pws}. An important one was to determine the interference of the $K^* K \bar K $ and $\rho \pi \pi$ loops, which could not be resolved in \cite{Ketzer:2015tqa}. Here we can do it, since we use the picture in which the 
$a_1(1260)$ is dynamically generated from the $K^* \bar K$ and $\rho \pi$ channels \cite{Roca:2005nm} and the theory provides the coupling of the resonance to these channels with a well defined relative sign. There are also other details concerning the regularization of the loops and their technical evaluation, which, in our case, we do using elements of the chiral unitary approach. Yet, the final results are very similar in what concerns the position, width and relative strength of the peak, giving a boost to the idea raised in \cite{Ketzer:2015tqa}, and providing a natural description of the peak seen in \cite{Adolph:2015pws} that, in view of this, cannot be accepted as a new resonance, once a conventional explanation for it has been found. On the other hand, the findings of the present work offer a remarkable example of how a decay mode of one resonance can peak at $200$ MeV higher energy than the nominal resonance mass.

\section{Formalism}
\label{formalism}
\begin{figure}[!ht]
\includegraphics[width=15cm,height=9cm]{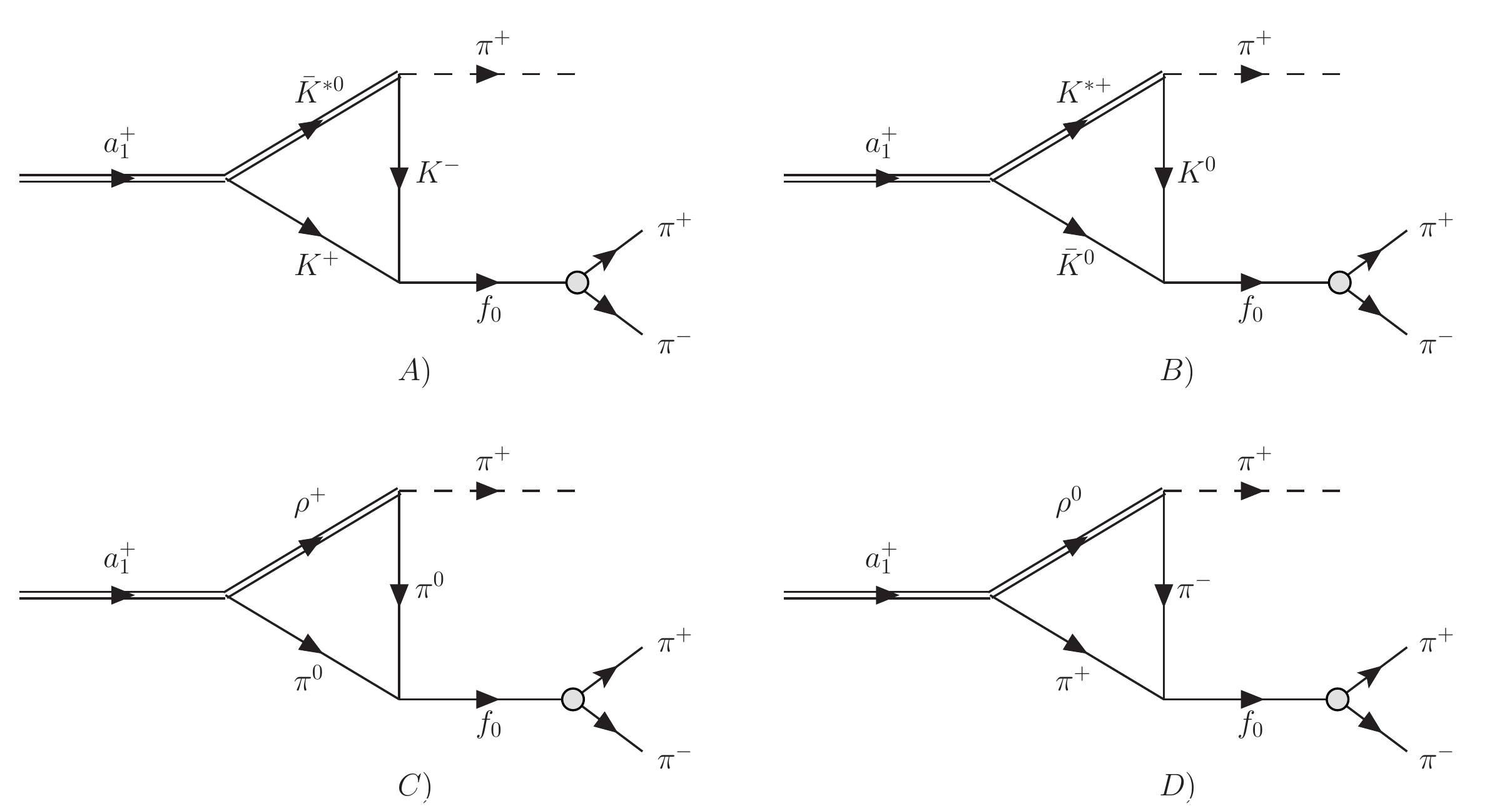}
\caption{Feynman diagrams for the process $a_1^+(1260)\rightarrow \pi^+\pi^+\pi^-$.}
\label{fig:diagrams}
\end{figure}

We want to evaluate the amplitude for the decay of the $a^+_1(1260)$ to $\pi^+\pi^+\pi^-$. As mentioned in the introduction, we will consider the $a_1(1260)$ as dynamically generated in a two coupled channel problem with building blocks $\rho\pi$ and $\bar{K}^*K$. Thanks to this assumption, the decay process observed in \cite{Adolph:2015pws} can be evaluated by means of the triangular mechanism shown in Fig. \ref{fig:diagrams}. 

In the four diagrams contributing to the process, the $a_1^⁺(1260)$ decays to the $\bar{K}^*K$ pair (diagrams $A)$ and $B)$) or to $\rho\pi$ (diagrams $C)$ and $D)$), followed by the decays of the $K^*$ to $\pi K$ and of the $\rho$ to $\pi\pi$. At this point, the two kaons or the two pions rescatter, leading to the $\pi\pi$ pair in the final state via the production of the $f_0(980)$ resonance. 

The production mechanism is completely analogous to the one already used to evaluate the decay $f_1(1285)\to a_0(980)\pi$ and we follow the same procedure. All the details of the calculation of the amplitude can be found in Ref. \cite{Aceti:2015zva}, while in this work we will only report the fundamental steps for convenience.

There are three vertices contributing to the diagrams. The $a_1PV$ vertex can be written as
\begin{equation}
\label{eq:vertex1}
-it_{1}= -ig_i\,C_1\epsilon_{a_1}^{\mu}\epsilon_{\mu}\, ,
\end{equation}
where $\epsilon_{a_1}$ is the polarization vector of the $a_1(1260)$ and $\epsilon$ the one of the $K^*$ for the diagrams $A)$ and $B)$ and of the $\rho$ for $C)$ and $D)$. The couplings $g_i$ of the $a_1(1260)$ to its building blocks, where $i=K^*\bar K,\,\rho\pi $, are obtained as the residue at the pole  of the scattering amplitude in $I=1$, which close to the pole can be written as
\begin{equation}
\label{eq:residues}
T_{ij}\simeq\frac{g_i g_j}{s-s_P}\, ,
\end{equation}
with $\sqrt{s_P}$ the position of the pole on the complex plane corresponding to the resonance. 

In Ref. \cite{Roca:2005nm}, the authors get the following values for the couplings in isospin base,
\begin{equation}
\begin{split}
g_{K^*\bar{K}}&=(1872-i1486)\,\textrm{MeV}\, ,\\
g_{\rho\pi }&=(-3795+i2330)\,\textrm{MeV}\, ,
\end{split}
\end{equation}
corresponding to a pole in $\sqrt{s_P}=(1011+i84)\ \textrm{MeV}$. We will use this same values throughout this entire work. However, the $a_1^+(1260)$ couples to the combinations with $I=1$, $C=+$ and $G=-$ of the $K^*\bar K$ and $\rho\pi$ pairs, which are represented by the states (for charge $+1$)
\begin{equation}
\begin{split}
\frac{1}{\sqrt{2}}(\bar{K}^*K-K^*\bar K)&=\frac{1}{\sqrt{2}}(\bar{K}^{*0}K^+-K^{*+}\bar{K}^0)\, ,\\
\rho\pi&=\frac{1}{\sqrt{2}}(\rho^0\pi^+-\rho^+\pi^0)\, .
\end{split}
\end{equation} 
The factor $C_1$ takes this into account and its values for the different diagrams of Fig. \ref{fig:diagrams} are listed in the second column of Table \ref{tab:factors}.

\begin{figure}[!ht]
\includegraphics[width=7cm]{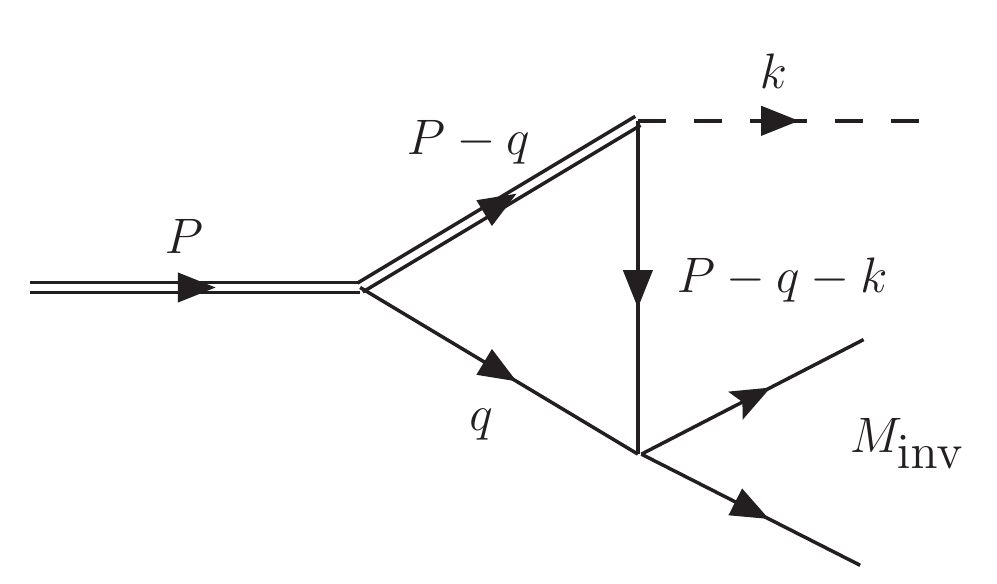}
\caption{Momenta assignment for the decay.}
\label{fig:momenta}
\end{figure}

The structure of the vertex for the $PPV$ interaction can be evaluated by means of the hidden gauge symmetry Lagrangian \cite{hidden1, hidden2, hidden3, hidden4}
\begin{equation}
\label{eq:lagrangianhg}
\mathcal{L}_{PPV}=-ig\ \langle V^\mu [P,\partial_\mu P]\rangle\ ,
\end{equation}
where the symbol $\langle\rangle$ stands for the trace in $SU(3)$ and $g=\frac{m_V}{2f}$, with $m_V\simeq m_{ \rho}$ and $f=93$\mev\ the pion decay constant. The matrices $P$ and $V$ in Eq. \eqref{eq:lagrangianhg} contain the nonet of the pseudoscalar mesons and the one of the vectors respectively. 
The resulting amplitude for the vertex can be written as
\begin{equation}
-it_{2}=-i\,g\,C_{2}\,(P-q-2k)_{\mu}\epsilon^{\mu}\ ,
\label{eq:vertex2}
\end{equation}
where the factors $C_2$ for the different diagrams are shown in the third column of Table \ref{tab:factors}. Fig. \ref{fig:momenta} shows the momenta assignment.

The third vertex corresponds to the mechanism for the production of the $\pi^+\pi^-$ pair in the final state, after the rescattering of the $K\bar{K}$ or $\pi\pi$, that dynamically generates the $f_0(980)$ resonance \cite{npa} as intermediate state. We will write the vertex as
\begin{equation}
\label{eq:vertex3}
-it_3=-it_{if}\ ,
\end{equation}
where $t_{if}$ is the $if$ element of the $5\times 5$ scattering matrix $t$ for the channels $\pi^+\pi^-$ (1), $\pi^0\pi^0$ (2), $K^+K^-$ (3), $K^0\bar{K}^0$ (4) and $\eta\eta$ (5) \cite{Liang:2014tia}. We have $i=1,2,3,4$ for the diagrams D), C), A) and B) respectively, while the index $f$ stands for channel $1$. The $t$ matrix is obtained using the Bethe-Salpeter equation, with the tree level potentials given in Refs. \cite{npa, Gamermann:2006nm} and compiled in Ref. \cite{Liang:2014tia}. The loop functions for the intermediate states are regularized using the cutoff method and the peak of the $f_0(980)$ is well reproduced using a cutoff of $630$\mev. We will need this parameter for the next steps of the calculation, being necessary in order to evaluate the loop integral in the diagrams of Fig. \ref{fig:diagrams}.

Using Eqs. \eqref{eq:vertex1}, \eqref{eq:vertex2} and \eqref{eq:vertex3}, we can write the expression of the amplitude for the four diagrams. As in Ref. \cite{Aceti:2015zva}, we will assume we are dealing with small three-momenta compared to the masses of the mesons, meaning that only the spatial components of the polarization vectors are different from zero. Hence, the completeness relation for the polarization vectors reduces to
\begin{equation}
\label{eq:completeness}
\sum_ {pol}\epsilon_{\mu}\epsilon_{\alpha}\simeq \sum_ {pol}\epsilon_{i}\epsilon_{j}=\delta_{ij}\ ;\ \ \ \ \ \ \mu=i,\ \ \alpha=j;\ \ i,j=1,2,3\ .
\end{equation}
The structure for the amplitude that we get is the same as in Ref. \cite{Aceti:2015zva}:
\begin{eqnarray}
\label{eq:amplabcd}
t^{A,B}&=&g_{\bar{K}^*K}\, g\, C\, \vec{\epsilon}_{a_1}\cdot\vec{k}\,(2I_1+I_2)\,t_{if}\, ,\nonumber\\
t^{C,D}&=&g_{\rho\pi}\, g\, C\, \vec{\epsilon}_{a_1}\cdot\vec{k}\,(2I_1^{\prime}+I_2^{\prime})\,t_{if}\, ,
\end{eqnarray} 
where the coefficients $C$ are simply given by the product of $C_1$ and $C_2$ and listed in the fourth column of Table \ref{tab:factors} and $I_1$, $I_2$, $I_1^{\prime}$ and $I_2^{\prime}$ are the loop integrals. We report here the expressions for the first two,
\begin{eqnarray}
I_1 &=& -\int \frac{d^3q}{(2\pi)^3} \frac{1}{8 \omega(q)
\omega^{\prime}(q) \omega^*(q)} \frac{1}{k^0 - \omega^{\prime}(q) -
\omega^{*}(q) + i \frac{\Gamma_{K^*}}{2}} \frac{1}{P^0 - \omega^*(q) - \omega(q) + i \frac{\Gamma_{K^*}}{2}} \nonumber \\
&& \times \frac{2P^0 \omega(q) + 2k^0 \omega^{\prime}(q) -
2(\omega(q) + \omega^{\prime}(q))(\omega(q)+\omega^{\prime}(q) +
\omega^{*}(q))}{(P^0 -
\omega(q)-\omega^{\prime}(q)-k^0+i\epsilon)(P^0+\omega(q)+\omega^{\prime}(q)-k^0-i\epsilon)}  ,\label{eq:loopintegral1}  \\
I_2 &=& -\int \frac{d^3q}{(2\pi)^3}
\frac{\vec{k}\cdot\vec{q}/|\vec{k}|^2}{8 \omega(q)
\omega^{\prime}(q) \omega^*(q)}
\frac{1}{k^0-\omega^{\prime}(q)-\omega^{*}(q) + i \frac{\Gamma_{K^*}}{2}}\,\frac{1}{P^0 - \omega^*(q) - \omega(q) + i \frac{\Gamma_{K^*}}{2}} \nonumber \\
&& \times \frac{2P^0 \omega(q) + 2k^0 \omega^{\prime}(q)-2(\omega(q)
+ \omega^{\prime}(q))(\omega(q) + \omega^{\prime}(q) +
\omega^{*}(q))}{(P^0-\omega(q) - \omega^{\prime}(q) -k^0 + i
\epsilon)(P^0 + \omega(q) + \omega^{\prime}(q) -k^0 -i \epsilon)}
,\label{eq:loopintegral2}
\end{eqnarray}
where $\omega(q)  = \sqrt{\vec{q}^{~2} + m^2_K}$, $\omega'(q) =\sqrt{(\vec q + \vec k)^2 + m^2_K}$, $\omega^*(q) =\sqrt{\vec{q}^{~2} + m^2_{K^*}}$ are the energies of the kaons and of the $K^*$ in the triangular loop. We included the finite width of the $K^*$ in its propagator, $\Gamma_{K^*}$, that we take equal to 48 MeV. The expressions for $I_1^{\prime}$ and $I_2^{\prime}$ are exactly the same with the substitutions $m_K\rightarrow m_{\pi}$, $m_{K^*}\rightarrow m_{\rho}$ and $\Gamma_{K^*}\to \Gamma_{\rho}$.

The upper limit for the numerical integrations in $d^3q$ of $I_1$, $I_2$, $I_1^{\prime}$ and $I_2^{\prime}$ is naturally provided by the chiral unitary approach, exactly as in Ref. \cite{Aceti:2015zva}, and it is given by the value of the cutoff for the loop function used in the meson-meson scattering to generate the $f_0(980)$, $q_{max}=630$ MeV. However, the integrals are already convergent without implementing $q_{max}$. The loops should also implement the cutoff used to generate the $a_1(1260)$, but its value is higher, around $1000$ MeV. Therefore, the use of $q_{max}=630$ MeV accounts for both cutoffs.
\begin{table}[ tp ]%
\begin{tabular}{c||c|c|c}
\hline %
Diagram &  $\phantom{-}\phantom{-}C_{1}\phantom{-}\phantom{-}$ & $\phantom{-}\phantom{-}C_{2}\phantom{-}\phantom{-}$ & $\phantom{-}\phantom{-}C\phantom{-}\phantom{-}$\\\toprule %
A) & $\phantom{-}\frac{1}{\sqrt{2}}$ & $-1$ & $-\frac{1}{\sqrt{2}}$ \\
B) & $-\frac{1}{\sqrt{2}}$ & $\phantom{-}1$ & $-\frac{1}{\sqrt{2}}$ \\
C) & $-\frac{1}{\sqrt{2}}$ & $-\sqrt{2}$ & $\phantom{-}1$ \\ 
D) & $\phantom{-}\frac{1}{\sqrt{2}}$ & $\phantom{-}\sqrt{2}$ & $\phantom{-}1$ \\\hline
\end{tabular}
\caption{Coefficients entering the evaluation of the triangular loop amplitudes of Fig. \ref{fig:diagrams}.}
\label{tab:factors}\centering %
\end{table}
We can sum the diagrams proceeding via the $\bar K^* K$ loop (A) and B)) and the ones proceeding via the $\pi \rho$ loop (C) and D)) obtaining, respectively,
\begin{equation}
\begin{split}
\label{eq:abandcd}
&t_{\bar{K}^*K}=-g_{\bar{K}^*K}\,g\,\frac{1}{\sqrt{2}}\,\vec{\epsilon_{a_1}}\cdot\vec{k}\,(2I_1+I_2)\,(t_{31}+t_{41})=\tilde{t}_{\bar{K}^*K}\,\vec{\epsilon_{a_1}}\cdot\vec{k}\, ,\\
&t_{\rho\pi}=g_{\rho\pi}\,g\,\vec{\epsilon_{a_1}}\cdot\vec{k}\,(2I_1^{\prime}+I_2^{\prime})\,(t_{11}+\sqrt{2}t_{21})=\tilde{t}_{\rho\pi}\,\vec{\epsilon_{a_1}}\cdot\vec{k}\, , 
\end{split}
\end{equation}
with the total amplitude for the decay given by
\begin{equation}
\label{eq:total}
t_{\textrm{tot}}=t_{\bar K^*K}+t_{\rho\pi}\, .
\end{equation}
The factor $\sqrt{2}$ multiplying $t_{21}$ in $t_{\rho\pi}$ takes into account that in the evaluation of the $t_{if}$ amplitudes the normalization $\frac{1}{\sqrt{2}}|\pi^0\pi^0\rangle$, appropriate for identical particles, is used and the good normalization must be restored.

We will use Eqs. \eqref{eq:abandcd} and \eqref{eq:total} to evaluate the invariant mass distributions of the process and its decay width.

\section{Results}
\begin{figure}[!ht]
\includegraphics[width=12cm]{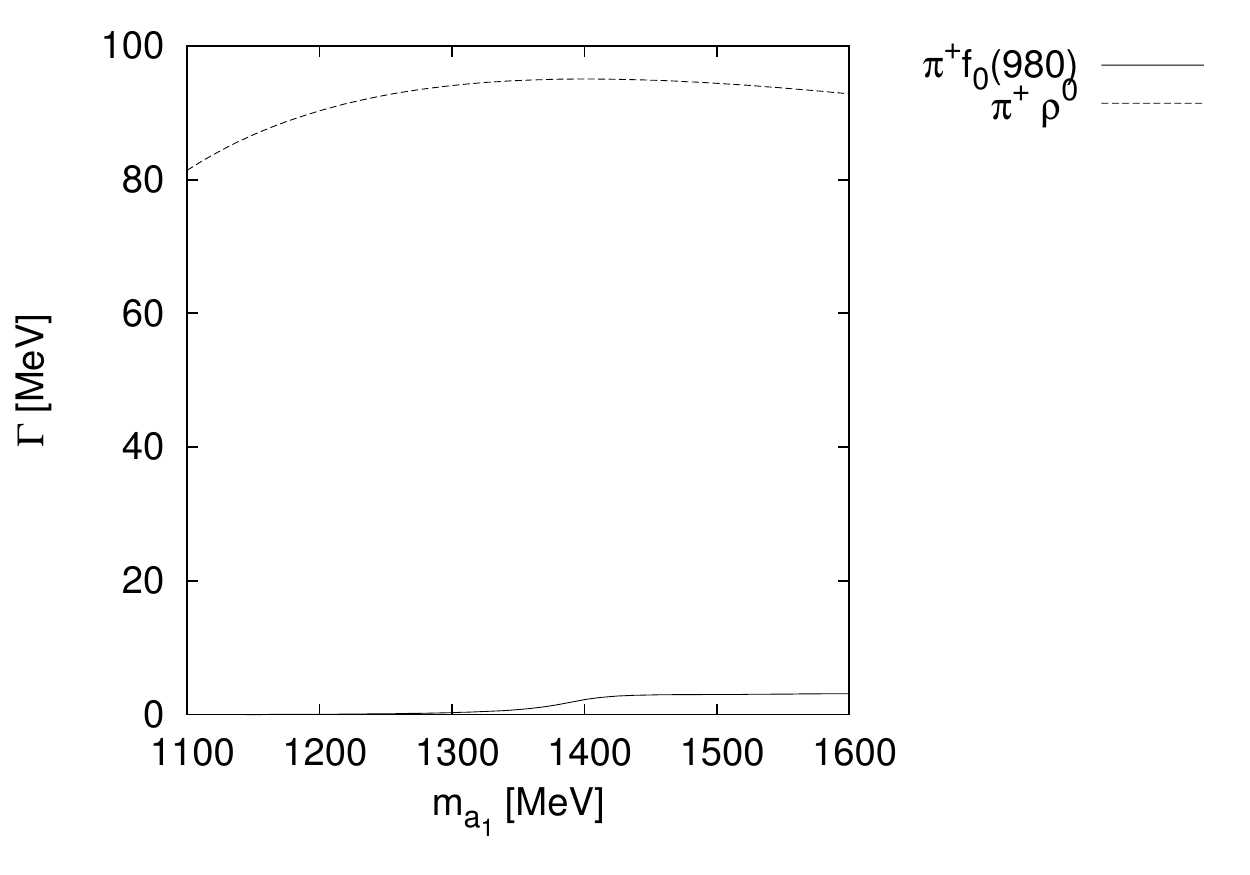}
\caption{Decay widths for the processes $a_1^+(1260)\to\pi^+f_0(980)$ (solid line) and $a_1^+(1260)\to \pi^+\rho^0$ (dashed line) as a function of the mass of the $a_1^+$.}
\label{fig:widths}
\end{figure}

The invariant mass distribution for the process is given by
\begin{equation}
\label{eq:invdistr}
\frac{d\Gamma}{dM_{\textrm{inv}}}=\frac{1}{(2\pi)^3}\frac{1}{4m_{a_1}^2}\frac{1}{3}|\vec{k}\,|^3 p_{\pi}|\tilde{t}|^2\, ,
\end{equation}
where $M_{\textrm{inv}}$ is the invariant mass of the final $\pi^+\pi^-$ pair. We omitted the sub-index for $\tilde t$ implying that it can be $\tilde t_{\bar K^*K}$, $\tilde t_{\rho\pi}$ or $\tilde t_{\textrm{tot}}$. The momenta in Eq. \eqref{eq:invdistr} are defined as
\begin{equation}
\begin{split}
\label{eq:momenta}
&p_{\pi}=\frac{\lambda^{1/2}(M_{\textrm{inv}}^2,m_{\pi}^2,m_{\pi}^2)}{2M_{\textrm{inv}}}\, ,\\
&|\vec{k}\,|=\frac{\lambda^{1/2}(m_{a_1}^2,m_{\pi}^2,M_{\textrm{inv}}^2)}{2m_{a_1}}\, .
\end{split}
\end{equation}
which are the momentum of the pion of the interacting pion pair in the $\pi^+\pi^-$ rest frame and the momentum of the spectator $\pi^+$ in the $a_{1}(1260)$ rest frame, respectively.
\begin{figure}[!ht]
\includegraphics[width=12cm]{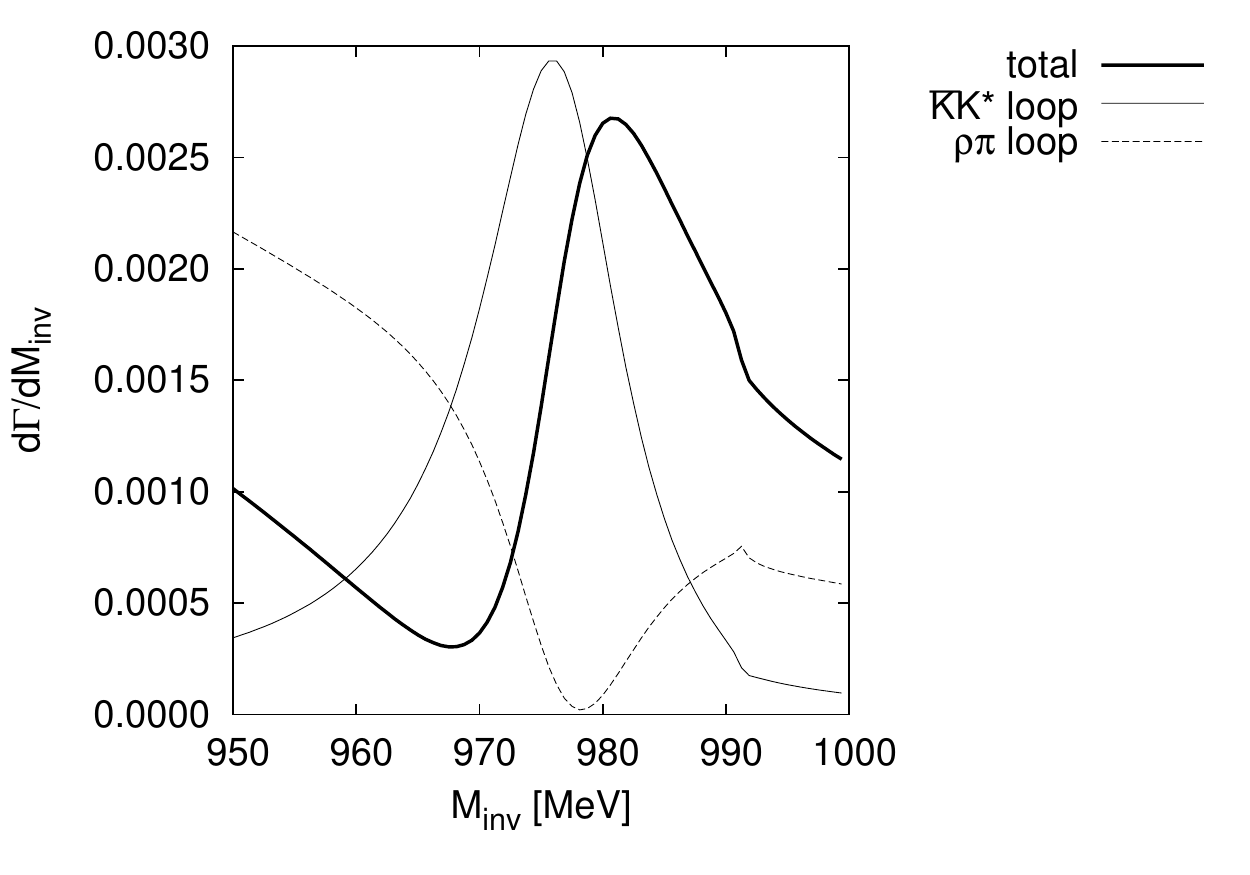}
\caption{$d\Gamma/dM_{\textrm{inv}}$ for $a_1^+(1260)\to\pi^+\pi^+\pi^-$ decay as a function of $M_{\textrm{inv}}$ in the region of the $f_0(980)$ considering only the $\bar K^*K$ loop diagrams (solid line), only the $\rho\pi$ loop diagrams (dashed line) and all the contributions (thick line). The widths of the $K^*$ and $\rho$ are removed in Eqs. \eqref{eq:loopintegral1} and \eqref{eq:loopintegral2}.} 
\label{fig:distr1230}
\end{figure}
\begin{figure}[!ht]
\includegraphics[width=16cm]{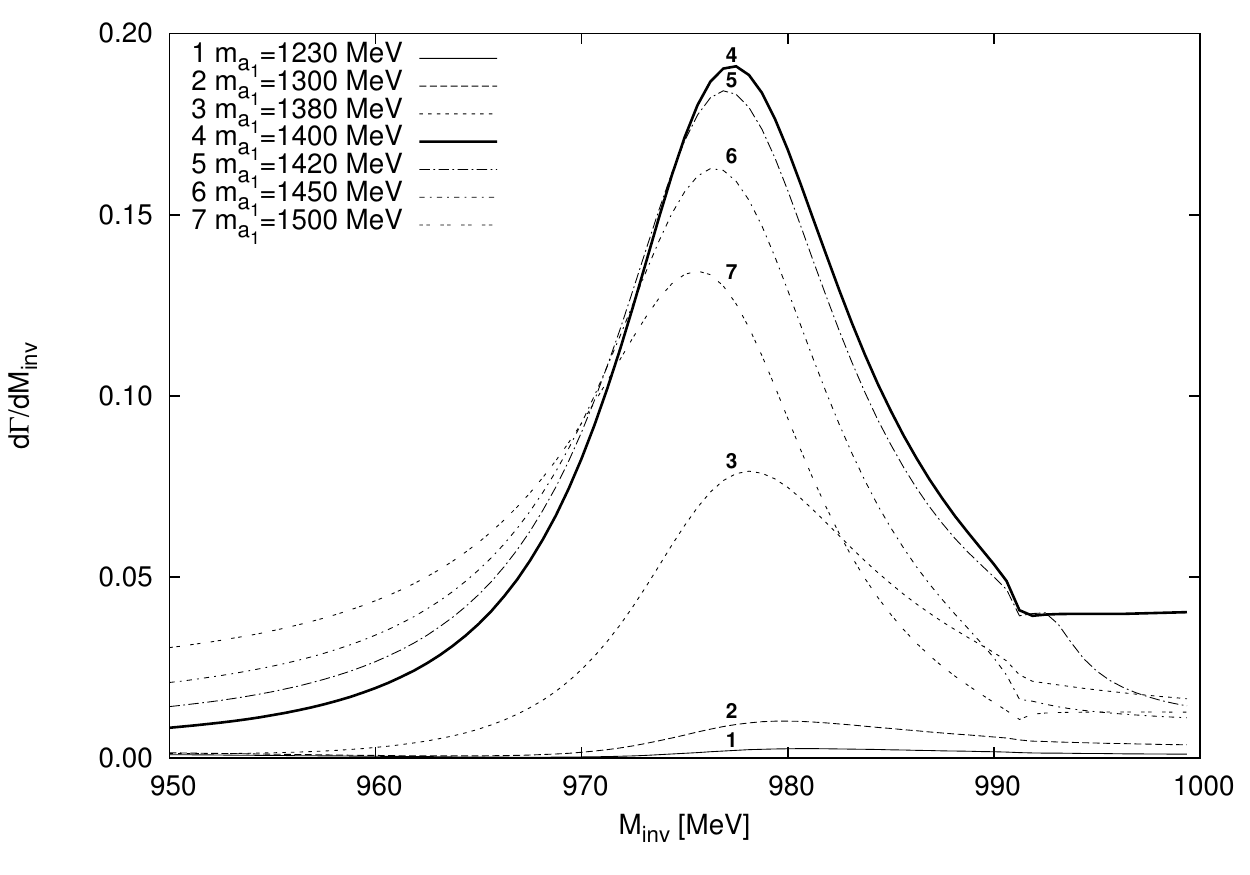}
\caption{$d\Gamma/dM_{\textrm{inv}}$ for $a_1^+(1260)\to\pi^+\pi^+\pi^-$ considering the contributions coming from the four diagrams of Fig. \ref{fig:diagrams} for different values of the mass of the $a_1^+$. The widths of the $K^*$ and $\rho$ in Eqs. \eqref{eq:loopintegral1} and \eqref{eq:loopintegral2} are not taken into account.}
\label{fig:distr}
\end{figure}
We plot $d\Gamma/dM_{\textrm{inv}}$ as a function of $M_{\textrm{inv}}$ in the region of the $f_{0}(980)$ for $\tilde t_{\textrm{tot}}$ (thick line), $\tilde t_{\bar K^* K}$ (solid line) and $\tilde t_{\rho\pi}$ (dashed line) in Fig. \ref{fig:distr1230} using for the $a_{1}^+$ resonance the value of the mass reported in the PDG \cite{Agashe:2014kda}, that is $1230$ MeV. Changing the mass of the $a_{1}^+$, leads to a variation of the strength at the peak of the distributions. As it is shown in Fig. \ref{fig:distr}, where $d\Gamma/dM_{\textrm{inv}}$ is plotted for values of $m_{a_1}$ up to $1500$ MeV, the strength at the peak reaches its maximum when the mass of the $a_1^+$ is around $1400-1420$ MeV. This calculation is done ignoring the widths of the $K^*$ and $\rho$ to show the effect of the singularity, peaking around $1420$ MeV. When the widths are kept, the effect of the singularity is softened as can be seen in Fig. \ref{fig:widths}.

Integrating in the invariant mass Eq. \eqref{eq:invdistr} with $\tilde t=\tilde t_{\textrm{tot}}$, we can evaluate the width for the decay $a_1^+(1260)\to \pi^+f_0(980)$ taking into account that 
\begin{equation}
\Gamma_{\pi^+f_0(980)}=\frac{3}{2}\,\Gamma_{\pi^+f_0(980)(f_0\to\pi^+\pi^-)}\, .
\end{equation}

In order to relate our findings to the results of Ref. \cite{Ketzer:2015tqa}, we also evaluate the width of the decay of the $a_1(1260)$ to its dominant decay channel $\rho\pi$, taking into account that
\begin{equation}
\Gamma_{\pi^+\rho^0}=\frac{1}{2}\Gamma_{\pi\rho}\, .
\end{equation}
The width $\Gamma_{\pi\rho}$ is given by the formula
\begin{equation}
\Gamma_{\pi\rho}=\frac{1}{8\pi}\frac{1}{m_{a_1}^2}\,g_{\rho\pi}^2\,\tilde p_{\rho}\, .
\end{equation}
where
\begin{equation}
\tilde p_{\rho}=\frac{\lambda^{1/2}(m_{a_1}^2, m_{\pi}^2, m_{\rho})}{2m_{a_1}}\, .
\end{equation}
The widths for both processes are evaluated as a function of the mass of the $a_1^+(1260)$ and we plot them in Fig. \ref{fig:widths}, taking into account the widths of the $\rho$ and $K^*$ in the loops of Eqs. \eqref{eq:loopintegral1} and \eqref{eq:loopintegral2}.

\begin{figure}[!ht]
\includegraphics[width=7cm]{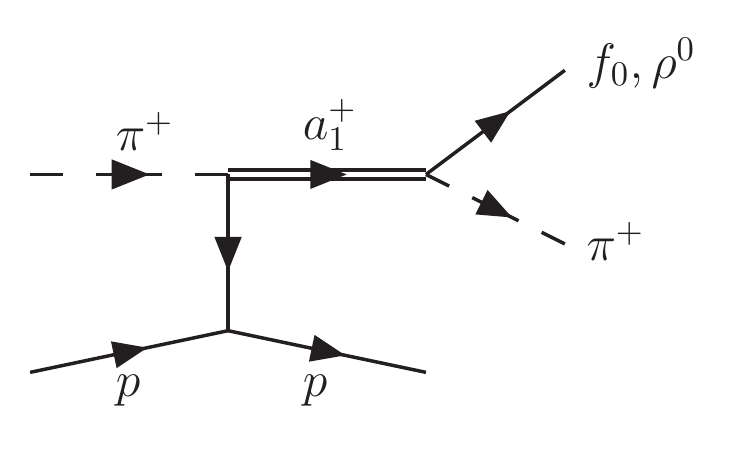}
\caption{Production of the $a_1(1260)$ by scattering of a high energy pion off a proton.}
\label{fig:full}
\end{figure}

\begin{figure}[!ht]
\includegraphics[width=12cm]{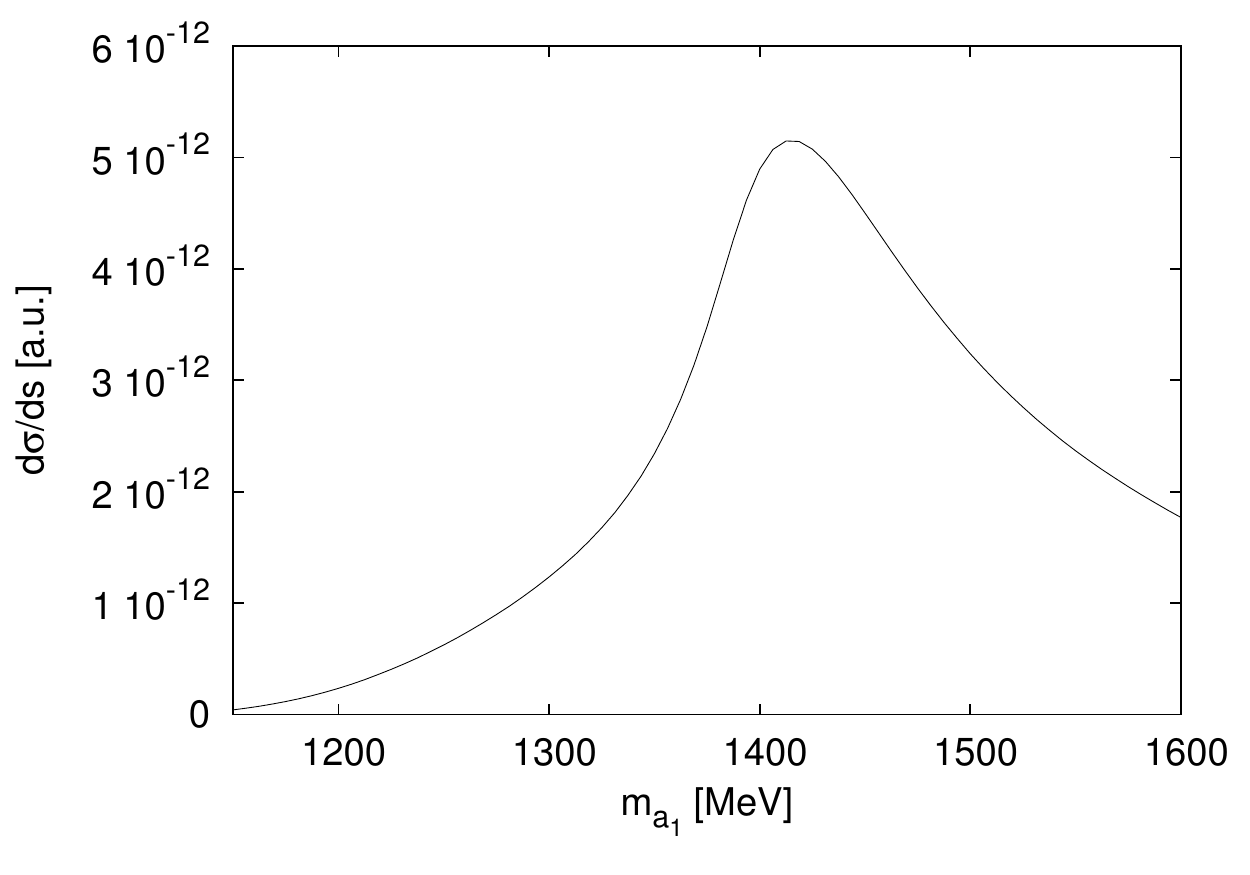}
\caption{Cross section for the decay $a_1^+(1260)\to\pi^+f_0(980)$ as a function of the center of mass energy.}
\label{fig:crossf0}
\end{figure}

\begin{figure}[!ht]
\includegraphics[width=12cm]{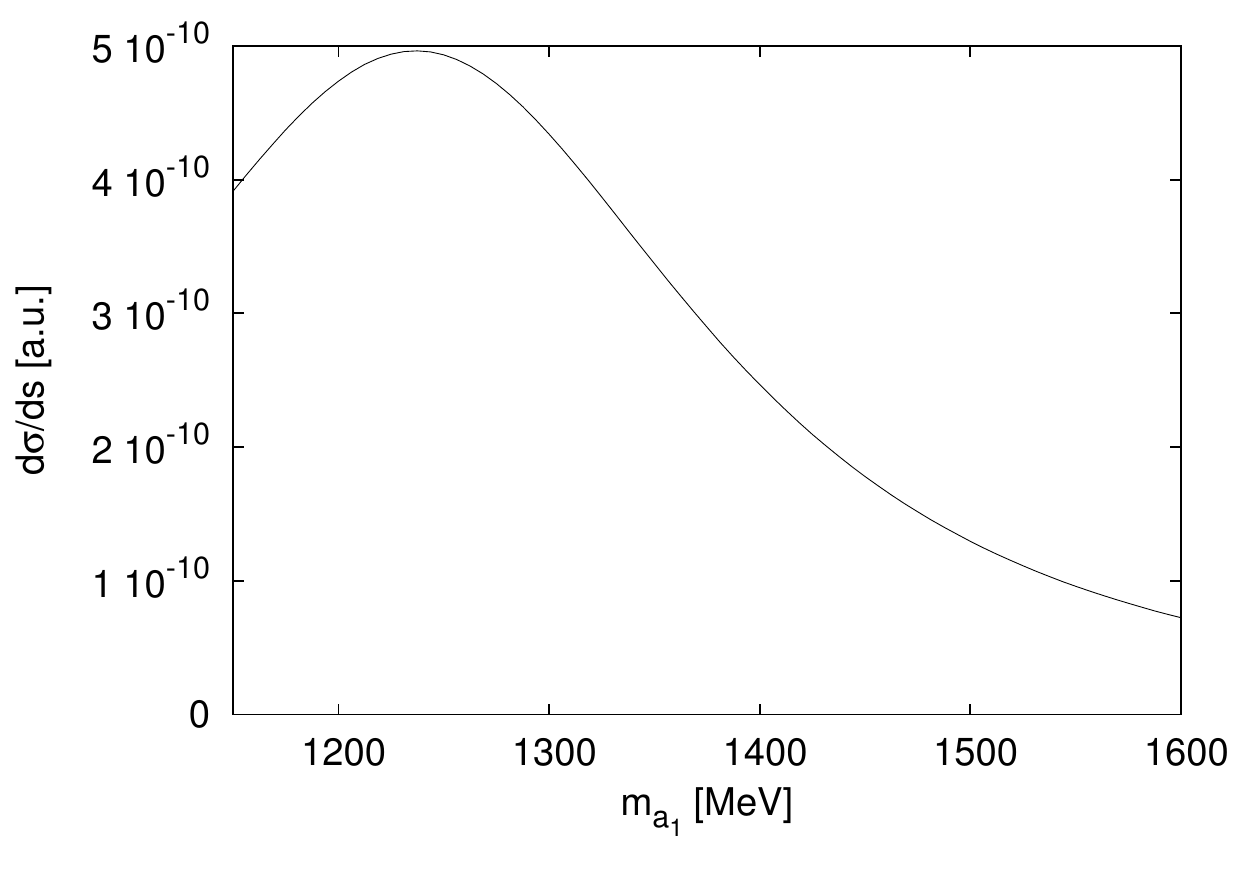}
\caption{Cross section for the decay $a_1^+(1260)\to\pi^+\rho^0$ as a function of the center of mass energy.}
\label{fig:crossrho}
\end{figure}

At this point, we can evaluate the cross sections of the diagrams for the production of the $a_1(1260)$ decaying to $\pi^+f_0(980)$ shown in Fig. \ref{fig:full}, which can be obtained, up to a constant factor, by multiplying the decay widths $\Gamma_{\pi^+ f_0(980)}$ and $\Gamma_{\pi^+\rho^0}$ by the square of the propagator of the $a_1(1260)$, such that
\begin{equation}
\label{eq:cross}
\frac{d\sigma_{\pi^+X}}{ds}\propto\frac{1}{(s-m^{*2}_{a1})^2+(m_{a1}^*\Gamma_{a_1})^2}\,\Gamma_{\pi^+X}(s)\, ,
\end{equation}
where $\sqrt{s}=m_{a_1}$ is the center of mass energy of the decay, $m_{a_1}^*=1230$ MeV the nominal mass of the $a_1(1260)$ and $\Gamma_{a_1}$ its width, that we take equal to $350$ MeV. The sub-index $X$ in Eq. \eqref{eq:cross} stands for the $f_{0}(980)$ or the $\rho^0$ depending on the process considered. In the case of the decay to $\pi^+f_0(980)$, which proceeds via $p$-wave ($\vec{\epsilon}_{a_1}\,\cdot\vec{k}$ in Eq. \eqref{eq:abandcd}) we also multiply by the Blatt-Weisskopf correction factor for $L=1$
\begin{equation}
\frac{1}{\sqrt{1+(|\vec{k}\,|\,R)^2}}\, ,
\end{equation}
with $R=0.25$ fm \cite{Manley:1984jz}. We plot the cross sections for the two cases in Figs. \ref{fig:crossf0} and \ref{fig:crossrho} respectively.


\section{Discussion of the results}
It is interesting to comment on these results. In the first place, the use of the chiral unitary approach to produce the $a_1(1260)$ as dynamically  generated from the $\bar K^* K$ and $\rho\pi$ channels allows one to obtain its couplings to $\bar K^* K$ and $\rho\pi$ with no ambiguity up to a global, irrelevant sign. Indeed, the coupled channel approach makes possible to obtain $T_{ij}$, which close to the peak behaves as Eq. \eqref{eq:residues}, providing the couplings. From $T_{11}$, for instance, one can obtain $g_{1}^2$, with an ambiguity in the sign of $g_1$. However, once this is fixed, the other couplings are obtained using $T_{1j}$ such that $g_j/g_1=T_{1j}/T_{11}$ and all the relative signs of the couplings, which are what matters in the evaluation of the loops, are unambiguously determined. This is different from  the work of Ref. \cite{Ketzer:2015tqa}, where the relative sign of $\rho\pi$ to $\bar{K}^*K$ could not be fixed. The mixture of the two channels in the $a_1{1260}\to\pi f_0(980)$ decay is relevant, as one can see in Fig. \ref{fig:distr1230}, although, as claimed in Ref. \cite{Ketzer:2015tqa}, the $\bar K^* K$ channel is the most important. The way to regularize the loops is also different than in Ref. \cite{Ketzer:2015tqa} and we rely upon the chiral unitary approach, as we have explained before, to control the large $\vec{q}$ contribution in the loop integrals. 

In spite of these differences, the results in Ref. \cite{Ketzer:2015tqa} are very similar in what concerns the shapes of the distributions and the relative weight of the $\pi^+f_0(980)$ and $\pi^+\rho^0$ decay modes, which comes to give support to the conclusions reached here.

In Fig. \ref{fig:crossrho} we can see the shape of the cross section for the $\pi^+p\to p\pi^+\rho^0$ reaction as a function of the invariant mass of the $\pi^+\rho^0$ system. We can see clearly the shape of the $a_1(1260)$ resonance, peaking around $1230$ MeV, and with its standard width. 

However, the same process producing $\pi^+ f_{0}(980)$, the $\pi^+p\to p\pi^+f_0(980)$ reaction,  has a very different shape, as shown in Fig. \ref{fig:crossf0}. We see a peak around $\sqrt{s}=1420$ MeV as in the COMPASS experiment \cite{Adolph:2015pws}, with a width  of about $150$ MeV, as also observed experimentally. This peak at $1420$ MeV, as in Ref. \cite{Ketzer:2015tqa}, is due to the triangular singularity of the diagrams A) and B) of Fig. \ref{fig:diagrams}. This occurs when the particles inside the loops are all placed on shell and, on top of it, the momenta are parallel or antiparallel. We have checked that the singularity occurs around $\sqrt{s}=1420$ MeV, and the momentum of the $K^*$, coming from the $a_1(1260)$ decay at rest, is parallel to the one of the the $\pi^+$. The inclusion of the width of the $K^*$ in the loop has a smoothing effect on the signal, and this is in line with what was also concluded by the authors of Ref. \cite{Achasov:2015uua} in the case of the $\eta(1405)\to \pi^0f_0(980)$ reaction.

It is also interesting to mention that our formalism differs technically from the one normally used in the study of the triangular singularities, where the integrations are performed using the Feynman parametrization. Instead, we do analytically the integration in $q^0$, while the $d^3q$ integration, which revert into two integrals, is done numerically. The results of the loop functions in Eqs. \eqref{eq:loopintegral1} and \eqref{eq:loopintegral2} are rewarding, showing explicitly the different cuts in terms of the explicit variables when pairs of interacting particles are placed on shell. The two physical cuts in our case correspond to having the $a_1$ decay into $\bar K^* K$ with the two particles on shell and the $K \bar K$ nearly on shell to produce the $f_0(980)$.

There is one more important result to comment. So far, the quantum number of the peak at $\sqrt{s}=1420$ MeV correspond to those suggested in Ref. \cite{Adolph:2015pws}, since we look at the decay of the $a_1(1260)$. In addition, the peak position and the width also agree with those seen in the experiment. The last magnitude to compare is the relative strength of $\pi^+\rho^0$ and $\pi^+f_0(980)$ production. Comparing Figs. \ref{fig:crossf0} and \ref{fig:crossrho}, we see that the ratio of the $\pi^+f_0(980)$ signal to that of $\pi^+f_0(980)$ at their respective peaks is of the order of $1\%$, as also found in \cite{Ketzer:2015tqa}, and in good agreement with experimental results \cite{Adolph:2015tqa}.

As we can see, the features observed in the experiment \cite{Adolph:2015pws,Adolph:2015tqa} are nicely reproduced by the decay mode of the $a_1^+(1260)$ into $\pi^+f_0(980)$, which proceeds via a triangular loop that has a singularity around $\sqrt{s}=1420$ MeV and is responsible for the peak seen at this energy. We should note that the decay mode $\pi^+\rho^0$ does not show any enhancement in the $1420$ MeV  energy region, a feature also observed experimentally. This last feature is relevant. Indeed, one peculiar thing of the triangular singularity is that the peak appears only in the particular reaction studied, the $a_1(1260)\to \pi f_0(980)$ in the present case. Failure to see the peak in other reactions gives extra support to the explanation given here for the experimental peak of \cite{Adolph:2015pws}.
\section{Conclusions}
We have studied the decay mode of the $a_1^+(1260)$ into $\pi^+f_0(980)$ and $\pi^+\rho^0$. The production mode of the $\pi^+\rho^0$ pair is obtained from the chiral unitary approach that generates this resonance from the interaction of the coupled channels $\bar K^* K-cc$ and $\pi\rho$, which allows to determine the couplings of the resonance to these channels. Also the $\pi^+f_0(980)$ production makes use of the chiral unitary approach for the dynamical generation of the $f_0(980)$ from the interaction of the $\pi\pi$, $K\bar K$ and $\eta\eta$ channels. In the present case, the $a_1(1260)$ decays to $\bar K^* K$ or $\pi\rho$, the $K^*$ decays to $K\pi$ and the $\rho$ to $\pi\pi$. After that, the resulting $K\bar K$ or $\pi\pi$ rescatter to produce the $f_0(980)$, which decays into the observed $\pi^+\pi^-$ channel. This decay mode is special since the triangle loop develops a triangular singularity around and energy of $1420$ MeV for the original $a_1$ state, which is produced in $\pi p$ reactions at high energies. We have shown that the position of the peak, its width and the relative weight of the $\pi^+f_0(980)$ production mode relative to the $\pi^+\rho^0$ one are all in agreement with the experimental findings. We could show that all these magnitudes appeared without the use of any free parameter extra to those inherent in the chiral unitary approach and, hence, they are absolute predictions of the theory. Having found a natural explanation for a peak based on known facts, our conclusion is that the ``$a_1(1420)$'' peak does not correspond to a new resonance but is just the manifestation of the decay mode of the well known $a_1(1260)$ resonance into the $\pi f_0(980)$ mode. 

On the other hand, one should stress that the present finding is rather remarkable, in the sense that it produces a peak for a decay mode of the resonance at an energy about 200 MeV higher than the nominal mass of the resonance.  This is quite a unique finding, and it might shed light into other cases seen in the PDG, where masses of the resonances are found rather different depending on the decay mode studied.

\section*{Acknowledgments}
This work is partly supported by the Spanish Ministerio de Economia y Competitividad and European FEDER funds under the contract number FIS2011-28853-C02-01, and the Generalitat Valenciana in the program Prometeo II, 2014/068. We acknowledge the Spanish Excellence Network on Hadronic Physics FIS2014-57026-REDT for the support. This work is also partly supported by the National Natural Science Foundation of China under Grant Nos. 11575076, 11375080 and supported by Program for Liaoning Excellent Talents in University under Grant No LR2015032. We would like to thank M. Mikhasenko and B. Grube for useful discussions and valuable information.

\end{document}